# Optical tweezers microrheology maps the dynamics of strain-induced local inhomogeneities in entangled polymers

**Manas Khan**[1,2,*], **Kathryn Regan**[1] **and Rae M. Robertson-Anderson**[1,†]

[1]Department of Physics and Biophysics, University of San Diego, San Diego, CA 92110, USA
[2]Department of Physics, Indian Institute of Technology – Kanpur, Kanpur – 208016, India

**Abstract**

Optical tweezers microrheology (OTM) offers a powerful approach to probe the nonlinear response of complex soft matter systems, such as networks of entangled polymers, over wide-ranging spatiotemporal scales. OTM can also uniquely characterize the microstructural dynamics that lead to the intriguing nonlinear rheological properties that these systems exhibit. However, the strain in OTM measurements, applied by optically forcing a micro-probe through the material, induces network inhomogeneities in and around the strain path, and the resultant flow field complicates the measured response of the system. Through a robust set of custom-designed OTM protocols, coupled with modeling and analytical calculations, we characterize the time-varying inhomogeneity fields induced by OTM measurements. We show that post-strain homogenization does not interfere with the intrinsic stress relaxation dynamics of the system, rather it manifests as an independent component in the stress decay, even in highly nonlinear regimes such as with the microrheological-LAOS (mLAOS) protocols we introduce. Our specific results show that Rouse-like elastic retraction, rather than disentanglement and disengagement, dominates the nonlinear stress relaxation of entangled polymers at micro- and meso- scales. Thus, our study opens up possibilities of performing precision nonlinear microrheological measurements, such as mLAOS, on a wide range of complex macromolecular systems.

[*]Corresponding author
mkhan@iitk.ac.in
[†]Corresponding author
randerson@sandiego.edu

**Word Count**
Main text: 2470
Captions: 126+166+156 = 448
Fig 1: (150/0.82) + 20 = 201
Fig 2: (150/0.50) + 20 = 320
Fig 3: (150/0.52) + 20 = 308
**Total: 3747**

Complex fluids and soft materials, such as networks of entangled polymers, have been widely studied over the past several decades due to the intriguing physics they exhibit – especially in response to stress or strain. Entangled polymers have been of particular recent interest as the Nobel Prize winning theory to describe their dynamics, de Gennes' reptation model, falls short of accurately describing their stress response in the nonlinear regime or when subject to non-uniform flow fields. Entangled polymer systems are also ideal candidates for designing multifunctional materials, as they display intriguing scale-dependent viscoelastic properties that can be precisely tuned by the concentrations, sizes and topologies of the constituent polymers [1-6].

Microrheology [7-11], which can probe viscoelasticity over lengthscales from below the network mesh size to larger than the size of the constituent polymers, offers a valuable approach to study these complex systems. In particular, active microrheology methods, such as using optical tweezers to drive a microsphere through a material with user-controlled rates and distances [12-19], provides access to the microscopic origin of nonlinear properties [13-16,18,19] and strain-induced rearrangements of network microstructures [17-19]. However, the flow field produced in these experiments is much more complex than analogous bulk rheology measurements and can lead to transient local inhomogeneities [13,20-23], which complicate accurate evaluation of rheological properties. The potential for inhomogeneities can become prohibitively large in nonlinear regimes where the physics is the most intriguing and least understood [1,5,18,21,24-26]. These issues further pose a major obstacle to implementing a microrheological analog of large-amplitude-oscillatory-shear (LAOS) measurements that have been proven extremely effective in elucidating the nonlinear response of polymer systems [27-32].

Thus, while optical tweezers microrheology (OTM) offers a promising route for probing rheological properties of entangled polymers and other soft materials, there has yet to be an accurate theoretical description of the resulting local inhomogeneities, much less a straightforward protocol for incorporating the details of the time-dependent flow field into analysis of rheological parameters. Previous theoretical studies have been limited to near-linear regime perturbations; and did not describe post-strain homogenization dynamics nor decouple this effect from the intrinsic relaxation of the system [13,20].

Here, we present a novel OTM protocol, supported by a theoretical framework, to elucidate the effect of strain-induced local inhomogeneities in polymeric materials. We use entangled DNA as a model system [33,34] and perform microrheological-LAOS (mLAOS) and single-strain OTM measurements. We measure the temporal variation of the inhomogeneity field and couple it to the time-varying osmotic pressure experienced by the microbead, such that we decompose the contributions of homogenization dynamics and intrinsic stress-relaxation to the experimentally measured stress response. Using a simple Fickian model we estimate the polymer concentration equilibration dynamics, and in turn, its manifestation in stress decay. We not only provide a complete description of the strain-induced inhomogeneity and its effect on the measured stress response, but we also introduce a valuable OTM technique – mLAOS. Our results answer key questions regarding how entangled polymers distribute local stress and relax following nonlinear strain, and open the door for straightforward analysis of a wide range of OTM measurements to investigate the nonlinear response of soft matter systems.

For both single-strain and mLAOS measurements (Fig 1, Supplemental Materials (SM)), we optically drag a micro-probe of radius $a = 2.25$ μm through a 1 mg/ml solution of 38 μm linear DNA at a constant strain rate $\dot{\gamma} = 3v/\sqrt{2}\, a = 85$ s$^{-1}$ (see SM) [18,34]. We simultaneously measure the resulting stress response of the system $\sigma(t) = F(t)/\pi a^2$ [18] via the force experienced by the probe, $F(t) = k\Delta x(t)$, where $\Delta x(t)$ is the displacement of the bead from the trap center, and $k$ is the trap constant. For single-strain experiments, we drag the probe through six different strain path lengths $L_{SP}$ corresponding to maximum strains ($\gamma_{max} = 3L_{SP}/\sqrt{2}\, a$) of 9.4 to 37.7 [20], after which we halt bead motion (SM, Figs. 1, 2). For mLAOS experiments, we oscillate the probe through the same strain path for 10 cycles. We vary the oscillation amplitude $L_{SP}$ and wait time before retracing the path $\Delta t_w$ to achieve oscillation periods $\Delta t$ of 0.44 s, 0.89 s and 1.49 s (SM, Figs. 1, 3). As demonstrated in Fig. S1, the chosen strain rates and magnitudes are sufficient to induce a nonlinear stress response.

The probe is much larger than the tube diameter ($d_T \approx$ 680 nm) and mesh size ($\xi \approx$ 140 nm) of the system [33-36]; and $\dot{\gamma}$ is well above the intrinsic relaxation rates $\tau_R^{-1}$ and $\tau_D^{-1}$, where $\tau_R \approx$ 0.6 s is the Rouse time associated with elastic retraction of polymer coils and $\tau_D \approx$ 40 s is the disengagement time over which an entangled polymer reptates out of its confining entanglement tube (see SM) [34,37]. Thus, as the probe moves through the network, it drags DNA along with it, creating a local density inhomogeneity. The inhomogeneity field can be described by a 'build-up' region with higher density of entanglement segments in front of the probe, and a lower density 'depletion' zone in the strain path (Fig. 1a). Upon cessation of the stage motion, the Stokes' drag force $F_d = 6\pi\eta a v$ (primarily responsible for the displacement $\Delta x(t)$ of the probe) vanishes, and the probe returns to the trap center by moving through the build-up region as the system equilibrates (Fig. 1b). This equilibration process is governed by the relaxation dynamics of the polymers in the build-up region, and manifested in the decay of $\Delta x(t)$, or $\sigma(t) = k\Delta x(t)/(\pi a^2)$ (Fig. 2a).

As shown in Fig 2, $\sigma(t)$ curves for all $\gamma_{max}$ values fit well to a double-exponential decay function, $\sigma(t) = C_1 e^{-t/\tau_1} + C_2 e^{-t/\tau_2}$, demonstrating that two distinct processes contribute to the stress relaxation. The corresponding relaxation times, $\tau_1$ and $\tau_2$, exhibit distinct dependences on $\gamma_{max}$ (Fig. 2b inset). While the shorter relaxation time, $\tau_1$, increases linearly with $\gamma_{max}$, $\tau_2$ is independent of $\gamma_{max}$, with an average value of ~0.6 s that corresponds remarkably well to the predicted Rouse time $\tau_R$ (SM) [34]. The agreement of $\tau_2$ with $\tau_R$, along with its independence from $\gamma_{max}$, indicates that this contribution to the relaxation arises from intrinsic relaxation mechanisms available to the polymers. Conversely, the linear increase of $\tau_1$ with $\gamma_{max}$ suggests that this process is driven by equilibration of the local inhomogeneity, which should depend on the strain.

To estimate the stress decay due to equilibration of the local inhomogeneity, we develop analytical calculations based on a toy model that describes the relaxation of the system following strain (Fig. 2c). All model assumptions and approximations are described and validated in SM. We treat the shape of the build-up region as roughly cylindrical with length $L$ and radius $R$. $L$ obtains an optimal value (dependent on $\dot{\gamma}$) and remains unchanged for increasing $\gamma$, while the build-up region grows cross-sectionally (increasing $R$) as new polymers are pushed into the region from the strain path. Therefore, the continuity equation $a^2 L_{SP} c = R^2 L(c_{bu}^0 - c) \approx R^2 L c_{bu}^0$ describes the polymer concentration in the build-up region, $c_{bu}^0$ ($\gg c$), immediately following the strain ($t = 0$) in terms of the bulk concentration, $c$. Polymers in the build-up region diffuse over time such that the average concentration $c_{bu}(t)$ decays from $c_{bu} = c_{bu}^0$ to $c_{bu} = c$ as $t \to \infty$. Diffusion occurs mostly along the radial direction of the cylindrical build-up region and can be described by Fick's law $J = -D \frac{\partial c_{bu}}{\partial r} \approx D \frac{c_{bu} - c}{R}$, where $D$ is the DNA diffusion coefficient in the bulk. While the density gradient is continuous in the build-up region, for simplicity we consider only the average value, $c_{bu}$, that changes to the bulk concentration $c$ across the boundary of the build-up region over a lengthscale of order $R$. Thus, the rate of concentration decay in the build-up region follows the differential equation $\frac{dc_{bu}}{dt} \approx -\frac{2J}{R} \approx \frac{2D}{R^2}(c - c_{bu})$. The solution of this equation, along with the boundary conditions, gives $c_{bu}(t) \sim c + (c_{bu}^0 - c)e^{-\frac{2Dt}{R^2}}$. Thus, the average polymer concentration in the build-up region decays exponentially with a characteristic timescale $\tau \propto R^2$. For constant $\dot{\gamma}$ perturbations, $R^2$ goes as $L_{SP}$ and hence $\tau$ varies proportionally with $L_{SP}$ (and thus $\gamma_{max}$). What remains to be determined is how the time-varying concentration $c_{bu}(t)$ relates to the measured stress $\sigma(t)$.

At a finite time, $t$, due to the mismatch of concentrations across the probe, it experiences an osmotic pressure force $f_{os}(t)$ that is proportional to $c_{bu}(t)$, as the concentration in the depletion region is negligible. Hence, the force equation for the probe can be written as $k\Delta x(t) +$ $6\pi\eta a(d\Delta x/dt) + f_{os}(t) + f_e(t) = 0$, where $f_e$ is the elastic recovery force exerted by the strained DNA. Assuming that the drag coefficient ($6\pi\eta a$) does not change significantly with time as the build-up region relaxes, the solution of the above first-order differential equation can be given by $\Delta x(t) = e^{-kt/6\pi\eta a} \int dt' \left(f_{os}(t') + f_e(t')\right) e^{kt'/6\pi\eta a}$. As shown in the previous paragraph, the homogenization dynamics leads to exponential decay of $c_{bu}(t)$, and in turn $f_{os}(t)$, with relaxation time $\tau$ that increases linearly with $L_{SP}$ (recall $c_{bu}(t) \sim c + (c_{bu}^0 - c)e^{-2Dt/R^2}$ and $R^2 \sim L_{SP}$). The force $f_e(t)$ also decays exponentially with a characteristic time, $\tau_R$, that is intrinsic to the system [34]. Since the time-dependence of both force terms in the above equation are exponential, the integration can be done easily to get $\Delta x(t) = Ae^{-t/\tau} + Be^{-t/\tau_R}$, where $A$ and $B$ are constants. Thus, $\Delta x(t)$ and, consequently, $\sigma(t)$ follow double-exponential decay with characteristic timescales of $\tau$ (similar to $\tau_2$), and $\tau_R$ ($\approx \tau_1$) as is manifested in our experiments.

Our theoretical estimations demonstrate that the time-varying concentration inhomogeneity manifests in experimental measurements as a separate exponential stress decay with timescale $\tau = \tau_1$, and does not interfere with the intrinsic stress response of the system. A longer $L_{SP}$ induces more inhomogeneity, i.e. denser build-up region, which takes longer to homogenize, resulting in slower decay of the osmotic pressure force.

While the build-up region dictates the stress response in single-strain experiments, the inhomogeneity field in the strain path, i.e the depletion region, becomes crucial for mLAOS measurements. We focus our mLAOS analysis on the maximum value of stress reached in each cycle, $\sigma_{max}$, seen as peaks in the repeated stress profiles (Fig. 3a). For all $\Delta t$ values $\sigma_{max}$ decreases in successive cycles and the decay of $\sigma_{max}(t)$ fits well to a double-exponential function, $\sigma_{max}(t) = C_3 e^{-t/\tau_3} + C_4 e^{-t/\tau_4}$, where $\tau_3$ and $\tau_4$ are the two characteristic decay times, $t = 0$ denotes the moment of occurrence of the first stress peak, and $t = n\Delta t$ (where $n$ is the cycle number starting with $n = 0$ for the first strain path) (Fig. 3b). The shorter relaxation time, $\tau_3$, is independent of $\Delta t$, with an average value of ~0.6 s ($\approx \tau_R$), while $\tau_4$ increases linearly with $\Delta t$ (Fig. 3b inset).

The forces acting on the probe in these mLAOS measurements are described by the same force balance equation $k\Delta x(t) + 6\pi\eta a(d\Delta x/dt) + f_{os}(t) + f_e(t) = 0$, so the stress variation can again be described by $\sigma(t) = \frac{k}{\pi a^2} e^{-kt/6\pi\eta a} \int dt' \left(f_{os}(t') + f_e(t')\right) e^{kt'/6\pi\eta a}$. Similar to the single-strain case, the second term decays exponentially with a characteristic timescale similar to $\tau_R$, due to the elastic relaxation of the strained DNA. Conversely, the osmotic force decays linearly with the decreasing concentration gradient of DNA across the probe in subsequent cycles.

We use a toy model depicted in Fig. 3c (and further described and validated in SM) to derive an analytical expression for this temporal decay of inhomogeneity. In the course of each cycle the probe sweeps away DNA from the strain path, leaving it initially fully depleted. In a time interval $\Delta t$, before the probe passes through that location again, DNA molecules from the neighboring region of lateral dimension $\sqrt{D\Delta t}$ diffuse into the strain path but get swept away in the next strain cycle. Hence, the DNA concentration in the neighboring region, $c_n(t)$, decreases with time. At a finite time $t$, the inward diffusive flux from the bulk to the neighboring region can be given by $J_{in}(t) = -D\frac{\partial c}{\partial r} \approx 2D \frac{c - c_n(t)}{\sqrt{D\Delta t}}$. Similarly, the outward flux from the neighboring region to the depletion region is $J_{out}(t) \approx 2D \frac{c_n(t) - c_{sp}}{\sqrt{D\Delta t}}$, where we have taken a time-averaged value of the periodically varying concentration at the strain-path, $c_{sp}$, since $c_n(t)$ decreases over a timescale several times longer than $\Delta t$. Therefore, the net change in $c_n(t)$ is given by the continuity equation $\frac{dc_n}{dt} = \frac{1}{4\pi a \sqrt{D\Delta t}} \left[2\pi\left(a + \sqrt{D\Delta t}\right) J_{in}(t) - 2\pi a J_{out}(t)\right] = \frac{1}{a\Delta t}\left[\left(a + \sqrt{D\Delta t}\right) c + a c_{sp} - \left(2a + \sqrt{D\Delta t}\right) c_n\right]$. The solution of this first-order differential equation describes the time dependence of $c_n(t)$, with an exponentially decaying leading term that satisfies the boundary condition $c_n(t = 0) = c$ and a decay time $\tau'$ that is proportional to $\Delta t$.

Using this framework, we can describe the time-dependent inhomogeneity in the strain-path, $c_{sp}(t)$, by a similar differential equation, $\frac{dc_{sp}}{dt} + \frac{2D}{a^2} c_{sp} = \frac{2D}{a^2} c_n$, with boundary condition $c_{sp}(t = 0) = 0$ where $t = 0$ denotes when the probe has just passed (beginning of cycle). Solving this equation gives $c_{\text{sp}} = c_n(1 - e^{-2Dt/a^2})$, which attains a maximum value $c_n(1 - e^{-2D\Delta t/a^2})$ at the end of each cycle ($t = \Delta t$), and gives rise to the osmotic force $f_{\text{os}}(t)$. Hence, $f_{\text{os}}(t)$, and in turn, the first term in $\sigma(t)$ follows the same exponential decay in $t$ as that of $c_n(t)$, with a decay time $\tau' \propto \Delta t$.

Our calculations validate the experimental observation that the strain-induced time-varying inhomogeneity in the strain-path causes an additional exponential decay in the peak stress measured at each cycle in an mLAOS experiment with a decay time proportional to the oscillation period, as $\tau_4 = \tau' \propto \Delta t$. While the dynamics of the inhomogeneity field are faster at higher oscillation frequencies, the amplitude of the oscillation determines the degree of inhomogeneity. The larger the amplitude ($L_{SP}$), the more DNA the probe drags from the strain path (directly) and neighboring region (indirectly) to the build-up zone in each cycle, creating a stronger inhomogeneity field. Our analysis thus demonstrates how to effectively separate out the complex dynamics of the inhomogeneity field induced by OTM measurements from the intrinsic stress response of macromolecular systems.

Here, we couple a robust set of optical tweezers measurements with modeling and analytical calculations to demonstrate the formation of time-varying inhomogeneity fields created by OTM experiments, and elucidate their effect on the microrheological response of polymeric materials. Our specific results for entangled DNA show that nonlinear stress relaxation of entangled polymers at micro- and meso- scales is governed by Rouse-like elastic retraction, occurring over the classically predicted Rouse time $\tau_R$, rather than the slower and oft-assumed dominant mechanism of entanglement tube disengagement (occurring over $\tau_D$). More generally, we show that the generation of local concentration inhomogeneities in macromolecular systems is inevitable in the nonlinear regime – when strain rates are much faster than the system relaxation rate or the strain amplitude is much larger than the system correlation length – such as with the presented mLAOS protocols. However, we illustrate that post-strain homogenization does not interfere with the intrinsic stress relaxation dynamics of the system, rather it manifests as an independent component in the stress decay. This additional component is identified by a distinct decay time that varies linearly with the oscillation period (mLAOS) or maximum strain value (single-strain). Thus, our study opens up possibilities of performing nonlinear OTM measurements, such as mLAOS, on a wide range of complex macromolecular systems to probe their intriguing nonlinear and scale-dependent rheological properties.

This research was funded by an AFOSR Biomaterials Award (No. FA9550-17-1-0249) awarded to RMR-A and an Initiation Grant from IITK awarded to MK.

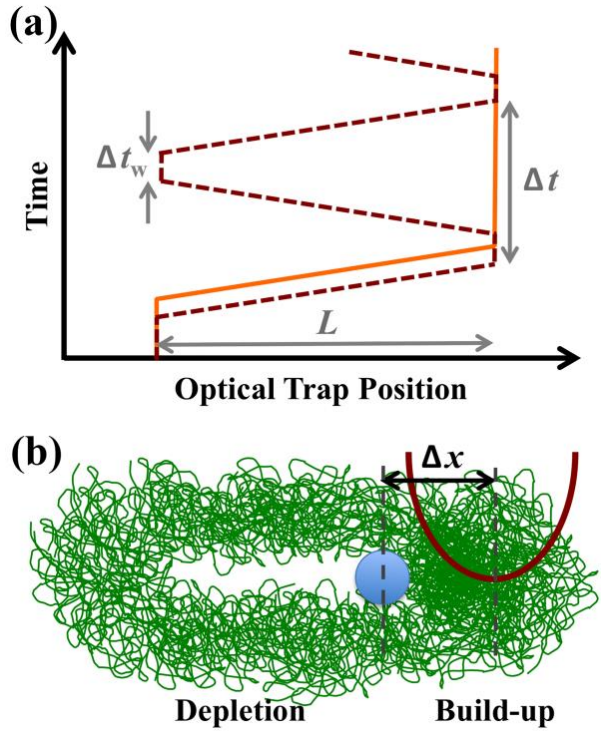

**Fig 1.** Schematic of OTM strain profiles (a) and resultant micro-strained state of entangled DNA (b). (a) Strain profiles represent the path of the trapped bead relative to the sample. In single-strain experiments (orange), the bead is dragged a distance $L_{SP}$ at a constant rate $\dot{\gamma}$, after which motion is halted. In mLAOS experiments (brown) a constant-rate oscillatory strain with period $\Delta t$ and wait-time $\Delta t_w$ is applied by repeatedly tracing the same strain path along forward and backward directions. (b) As the probe moves through the solution, DNA in the strain path accumulates, creating a higher density 'build-up' region in front of the bead, and lower density 'depletion' region behind it. At the same time, Stokes' drag displaces the bead from the trap center by a distance $\Delta x$.

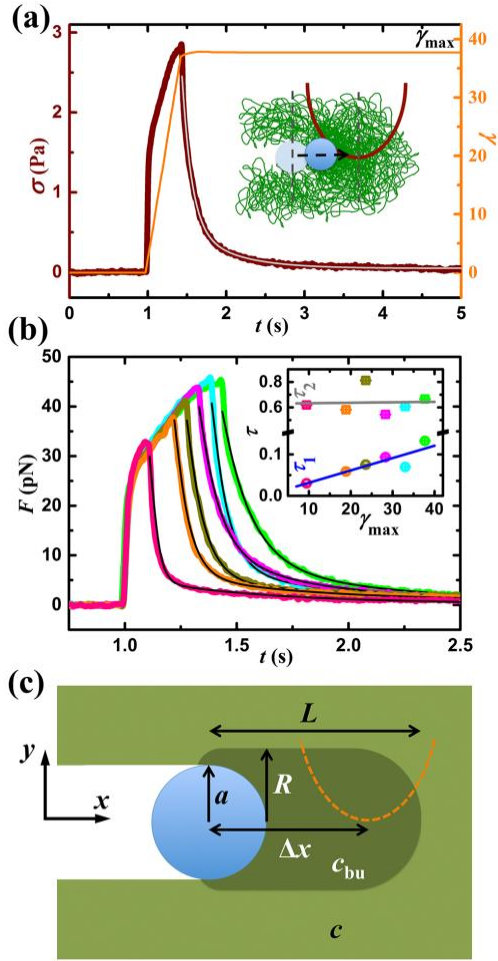

**Fig 2.** Stress relaxation of entangled DNA is mediated by Rouse-like elastic retraction coupled with homogenization of non-uniform concentration profiles. (a) Sample applied strain $\gamma(t)$ (orange line, right axis) and measured stress response $\sigma(t)$ (brown line, left axis) versus time $t$. The stress relaxation at constant strain $\gamma_{max}$ is fit to a double-exponential decay (grey line). Data shown is for $\gamma_{max} = 37.7$. Inset: Bead motion through the build-up region towards the trap center as the system relaxes. (b) $\sigma(t)$ curves for $\gamma_{max}$ values of 9.4 (pink), 18.9 (orange), 23.6 (olive), 28.3 (magenta), 33.0 (cyan), and 37.7 (green), all fit to double-exponential decays with decay times $\tau_1$ and $\tau_2$ (black lines). Inset: $\tau_1$ and $\tau_2$ (color-coded circles and squares with error bars), versus $\gamma_{max}$ with linear fits to the data. (c) Schematic of the model and associated variables: micro-probe radius ($a$), radius ($R$) and length ($L$) of 'build-up' region, probe displacement from trap center ($\Delta x$), and DNA concentrations in the build-up region ($c_{bu}$) and in bulk ($c$).

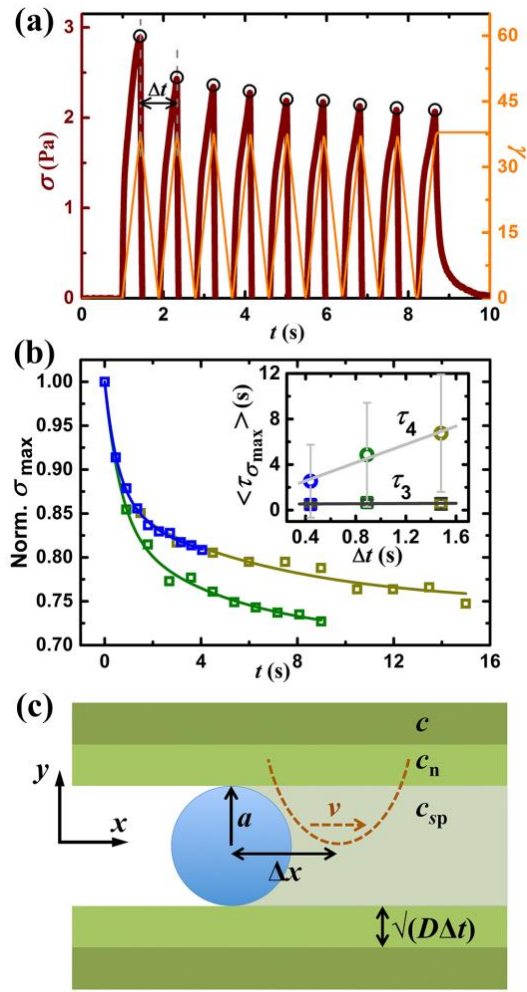

**Fig 3.** The maximum stress induced in entangled DNA during mLAOS measurements exhibits two-phase exponential decay. (a) Sample applied strain $\gamma(t)$ (orange, right axis) and measured stress $\sigma(t)$ (brown, left axis) during an mLAOS measurement with period $\Delta t$. Black circles denote the maximum stress measured for each oscillation $\sigma_{max}(t)$. (b) $\sigma_{max}(t)$ values (hollow squares) for ten oscillations with $\Delta t$ values of 0.44 s (blue), 0.89 s (green) and 1.49 s (olive), all fit to double-exponential decays (solid lines) with decay times $\tau_3$ and $\tau_4$. Inset: $\tau_3$ and $\tau_4$ (color-coded squares and circles with error bars) versus $\Delta t$ with linear fits to the data. (c) Schematic of the model with the associated variables: probe radius ($a$), probe speed ($v$), displacement of probe from the trap center ($\Delta x$), width of the neighboring region ($\sqrt{D\Delta t}$) that comprises DNA that diffuses into the strain path in wait time $\Delta t$, DNA concentrations in the strain path ($c_{sp}$), neighboring region ($c_n$), and bulk ($c$).

# Optical tweezers microrheology maps the dynamics of strain-induced local inhomogeneities in entangled polymers

Manas Khan[1,2,*], Kathryn Regan[1] and Rae M. Robertson-Anderson[1, †]

[1]Department of Physics and Biophysics, University of San Diego, San Diego, CA 92110, USA
[2]Department of Physics, Indian Institute of Technology – Kanpur, Kanpur – 208016, India

*Note: All References and Figures cited correspond to those in the main text*

**DNA Preparation and Characterization:**

Double-stranded 115 kbp (38 μm) DNA was prepared, as previously described [33], by replication of a cloned bacterial artificial chromosome construct in E. coli, followed by extraction, purification, and restriction enzyme treatment to convert supercoiled constructs to linear form. Linear DNA constructs were resuspended in aqueous buffer [10 mM Tris-HCl (pH 8), 1 mM EDTA, 10 mM NaCl] to a concentration of 1.0 mg/ml, corresponding to ~40$c^*$ [34] and ~7x the critical entanglement concentration $c_e$ [38]. The predicted entanglement tube diameter for this system is $d_T \approx$ 680 nm and the mesh size is $\xi \approx$ 140 nm [33-36]. The predicted Rouse time, which is the timescale over which deformed polymer coils elastically retract, is $\tau_R = 2R_G^2/\pi^2 D_G \approx$ 0.6 s, where $R_G$ and $D_G$ are the radius of gyration and diffusion coefficient in the dilute limit [34, 35, 37]. The predicted disengagement time, which is the timescale over which an entangled polymer reptates out of its confining entanglement tube and into a new tube, is $\tau_D = 18(R_G/d_T)^2 \tau_R \approx$ 40 s [34, 35, 37].

**OTM Measurement Details:**

For both single-strain and microrheological-LAOS (mLAOS) measurements depicted in Fig. 1, a trace amount of polystyrene microsphere probes with radius $a$ = 2.25 μm were embedded in a 1 mg/ml solution of 115 kbp (38 μm) solution of DNA. An isolated probe, located >50 μm (>20$a$) away from the surfaces of the sample cell, was optically dragged through the entangled DNA at a constant speed $v$ = 90 μm/s, corresponding to a strain rate $\dot{\gamma} = 3v/\sqrt{2}\,a = 85$ s$^{-1}$ [20], Weissenberg number $Wi = \dot{\gamma}\tau_d \approx 3400$ and Rouse Weissenberg number $Wi_R = \dot{\gamma}\tau_R \approx 51$. The optical trap used to perform both measurements has been fully described and validated previously [18].

For single-strain experiments, we drag the probe through different strain path lengths $L_{SP}$ = 10, 20, 25, 30, 35, and 40 μm, corresponding to maximum strain values of $\gamma_{max} = 3L_{SP}/\sqrt{2}\,a$ [20], after which we halt the bead motion (Fig. 1a). Thus, the applied strain increases from $\gamma = 0$ to $\gamma = \gamma_{max}$ after which it remains constant. For mLAOS experiments, we drag the probe forward and backward along the same strain path for 10 cycles. We vary the oscillation amplitude, $L_{SP}$, and the wait time before retracing the strain path, $\Delta t_w$, to achieve three different oscillation periods, $\Delta t$ = 0.44 s ($L_{SP}$ = 20 μm, $\Delta t_w$ = 0), 0.89 s ($L_{SP}$ = 40 μm, $\Delta t_w$ = 0) and 1.49 s ($L_{SP}$ = 40 μm, $\Delta t_w$ = 0.3 s) (Fig. 1a). For all OTM measurements, the applied strains corresponding to the $L_{SP}$ values are in the nonlinear regime as demonstrated in Fig. S1. For both types of measurements, the stress response of the system, $\sigma(t) = F(t)/\pi a^2$ [18], is measured via the force experienced by the probe, $F(t) = k\Delta x(t)$, where $\Delta x(t)$ is the displacement of the bead from the center of the trap, and $k$ is the trap constant. The instantaneous value of the trapping laser deflection, which is proportional to $\Delta x(t)$, is recorded at 20 kHz using a position sensing diode that converts the deflection of the forward scattered light to a voltage signal. By measuring the Stokes' drag in a viscous medium (50% (v/v) glycerol in water) the value of $k$ is determined as 83 pN/V.

**Theoretical Model and Calculation Details:**

To understand the relaxation dynamics of the strain-induced inhomogeneity field and the linear dependence of the relaxation times ($\tau_1$ and $\tau_4$) on the strain-profile, we have presented analytical estimations based on toy models, which by definition do use a few assumptions and simplifications. There are many hurdles to formulating a definitive theoretical description and model of the local strain-field and density-field in nonlinear microrheology measurements [20]. Hence, a few assumptions and simplifications are essential to attempt to unravel the dynamics of the inhomogeneity field created by the applied strain in our experiments. Here, we enumerate all of the important assumptions and simplifications we have used in our analytical calculations and justify their validity.

(a) *The shape of the build-up region*: We have assumed that the build-up region is cylindrical in shape. The length of the cylindrical build-up region depends solely on the applied strain rate, while the cross-sectional area varies with the final strain. This assumption is based on the observations reported in Ref. [18]. There, a probe is dragged through a similar DNA system as ours with optical tweezers and then the probe is released to measure the recoil force exerted by the deformed network. The strength of the recoil is determined by the amount of accumulated and deformed DNA molecules in front of the sphere at the end of the applied strain. This is nothing but the length of the build-up region where the deformed DNA molecules get accumulated as the probe sphere is dragged through the system. It was shown that the recoil varied with the applied strain rate and not with strain. Hence, we assume that the length of the build-up region is determined by the strain rate and not the strain. This justifies our assumption that as the strain is increased at a constant rate (thus keeping the length of the build-up region unchanged), the cross-sectional area of the build-up region would increase accordingly, as more DNA molecules accumulate in the build-up region.

(b) *The shape of the depletion zone*: As a probe sphere of diameter $d = 4.5$ μm is dragged through a network with mesh size (~140 nm) and tube diameter (~680 nm) much smaller than $d$, DNA molecules in the strain path are swept away leaving behind a depletion zone. At high shear rates, as in the case here, the length of the depletion zone is longer than at least a few probe diameters while the cross-section is comparable to $d$ [39]. DNA molecules from the neighboring region diffuse into the depletion zone to curtail the length of the depletion zone.

Hence, the steady-state length of the depletion region depends on the strain rate and the diffusion coefficient of the DNA molecules. Because of this, the exact shape of the depletion region is expected to be conical. Here, we have assumed the shape of the depletion zone to be cylindrical to estimate the diffusion of DNA molecules from the neighboring region to the depletion zone in a simplified way. However, the exact shape of the depletion zone does not affect our analytical calculation in any way, as our estimate considers only the concentration gradient across the probe (to calculate the force experienced by the probe due to osmotic pressure) and the concentration of DNA molecules in the depletion zone (that have diffused into the depletion zone from the neighboring region) after time-period $\Delta t$. Neither of these quantities depends on the exact shape of the tail of the depletion zone.

(c) *Continuity equations*: We have used continuity equations to determine the concentration of DNA molecules in and around the depletion zone and the build-up region. Since the number of DNA molecules in the system remains conserved, the local concentration inhomogeneity field created by the applied strain must obey continuity equations.

(d) *Fickian diffusion model*: We have used Fickian diffusion equations to describe the homogenization dynamics of the concentration-inhomogeneity field created by the applied strain. This assumption is valid if the DNA molecules in our system do not experience any restriction that forbids their normal relaxation or diffusion dynamics. This criterion is evident in previously reported studies on similar systems [34, 38], as well the results we present here, thereby justifying the application of a Fickian diffusion model in our analytical calculations.

(e) *Discretization of density gradient*: In steady state, both the build-up region and the depletion zone would have continuous boundaries with a smooth concentration gradient due to the homogenization dynamics governed by Fickian diffusion of the DNA molecules. However, for simplification, we have discretized the inhomogeneity field by replacing the continuous concentration gradients with regions of varying concentration. Although the regions have well-defined dimensions, determined by the continuity equation or the diffusion equation, the concentration changes over a finite length across the boundary. Thus, the discretization of the inhomogeneity field in our toy model is a very close approximation of the real continuously varying concentration gradients.

With these assumptions and simplifications our analytical calculations estimate the homogenization dynamics of the density-inhomogeneity field and establish that this process is indeed the origin of the relaxation times $\tau_1$ and $\tau_4$. Therefore, the analytical estimations that support our experimentally obtained conclusions are based entirely on validated assumptions and simplifications.

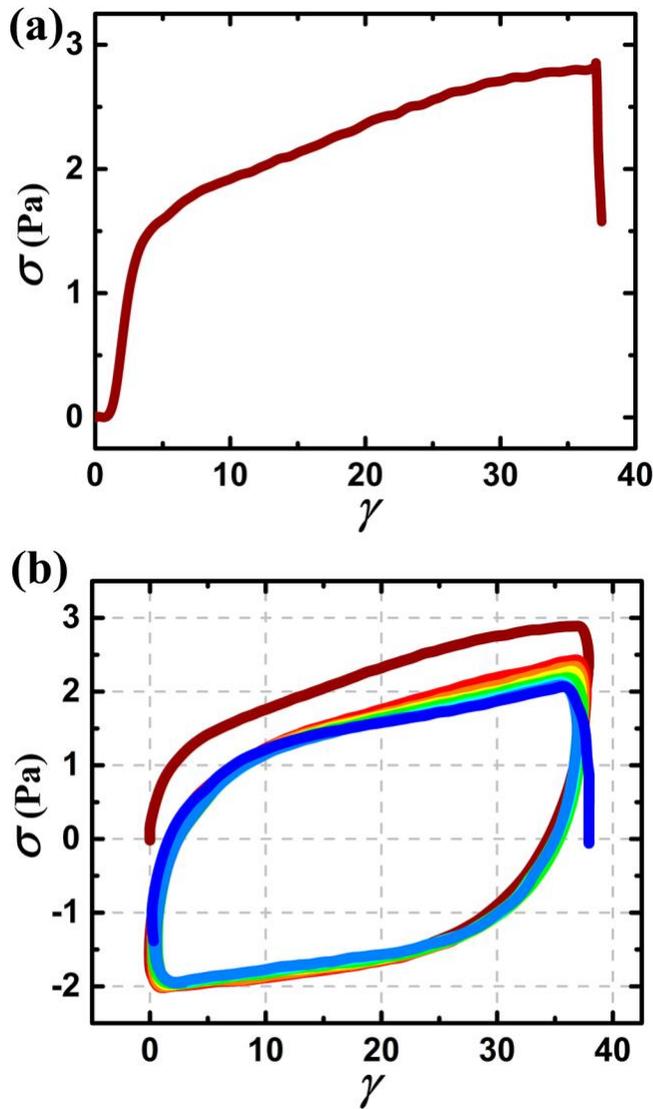

**Figure S1. Stress-strain profile for single-strain (a) and mLAOS (b) measurements.** (a) The stress response ($\sigma$) versus the applied strain ($\gamma$) starts exhibiting nonlinearity at around $\gamma = 5$. (b) The Lissajous plot shows nonlinear stress response ($\sigma$) versus the applied periodic strain ($\gamma$), where each strain-cycle is plotted with different color for clarity. Note that the stress peaks ($\sigma_{max}$) in the top right corner of the plot not only decay in successive strain cycles, but they also do not occur at the same strain value $\gamma$. Because $\gamma$ is periodic in time, the shift of $\sigma_{max}$ with respect to $\gamma$ indicates that the stress peaks do not appear at equal temporal spacing, as expected in the nonlinear regime. Further, note that the stress peaks are shifted from the strain peaks, also expected in the nonlinear regime.